\pdfoutput=1
\documentclass[journal,english,twocolumn]{IEEEtran}

\usepackage[T1]{fontenc}
\usepackage{amsmath}
\usepackage{graphicx}
\usepackage{enumerate}
\usepackage{amssymb}
\usepackage[dvipsnames]{xcolor}
\makeatletter
\usepackage{cite}
\usepackage[caption=false,font=footnotesize]{subfig}
\makeatother
\usepackage{babel}
\usepackage{siunitx}
\usepackage{tikz}
\usepackage[final]{changes}
\newcommand\copyrighttext{%
	\footnotesize \copyright 2020 IEEE. Personal use of this material is permitted. Permission from IEEE must be obtained for all other uses, in any current or future media, including reprinting/republishing this material for advertising or promotional purposes, creating new collective works, for resale or redistribution to servers or lists, or reuse of any copyrighted component of this work in other works.}
\newcommand\copyrightnotice{%
	\begin{tikzpicture}[remember picture,overlay]
	\node[anchor=south,yshift=5pt] at (current page.south) {\fbox{\parbox{\dimexpr\textwidth-\fboxsep-\fboxrule\relax}{\copyrighttext}}};
	\end{tikzpicture}%
}

\begin{document}
\bstctlcite{IEEEexample:BSTcontrol}

\title{Noise Temperature of Phased Array Radio Telescope: The Murchison Widefield Array and the Engineering Development Array}

\author{Daniel C. X. Ung, Marcin Sokolowski, Adrian T. Sutinjo, David B. Davidson

\thanks{IEEE Trans. Antennas. Propagat., accepted, 6 March 2020. The authors are with the International Centre for Radio Astronomy Research/Curtin Institute of Radio Astronomy, Curtin University, Bentley,
	WA 6102, Australia (e-mail: daniel.ung@icrar.org).}}




\maketitle
\copyrightnotice
\begin{abstract}
\added{This paper} presents a framework to compute the receiver noise temperature ($T_{\mathrm{rcv}}$) of two low-frequency radio telescopes, the Murchison Widefield Array (MWA) and the Engineering Development Array (EDA). The MWA was selected because it is the only operational low-frequency Square Kilometre Array (SKA) precursor at the Murchison Radio-astronomy Observatory, while the EDA was selected because it mimics the proposed SKA-Low station size and configuration. It will demonstrated that the use of an existing power wave based framework for noise characterization of multiport amplifiers is sufficiently general to evaluate $T_{\mathrm{rcv}}$ of phased arrays. \added{The calculation of $T_{\mathrm{rcv}}$ was done} using a combination of measured noise parameters of the low-noise amplifier (LNA) and simulated $S$-parameters of the arrays. The calculated values \added{were compared} to measured results obtained via astronomical observation and both results are found to be in agreement. Such verification is lacking in current literature. It was shown that the receiver noise temperatures of both arrays are lower when compared to a single isolated element. This is caused by the increase in mutual coupling within the array which is discussed in depth in this paper.
\end{abstract}

\begin{IEEEkeywords}
	Aperture arrays, Mutual coupling, Noise receiver temperature, Radio Telescope, Radio astronomy  
\end{IEEEkeywords}

\section{Introduction}
\begin{figure}[]
	\centering
	\includegraphics[width=\linewidth]{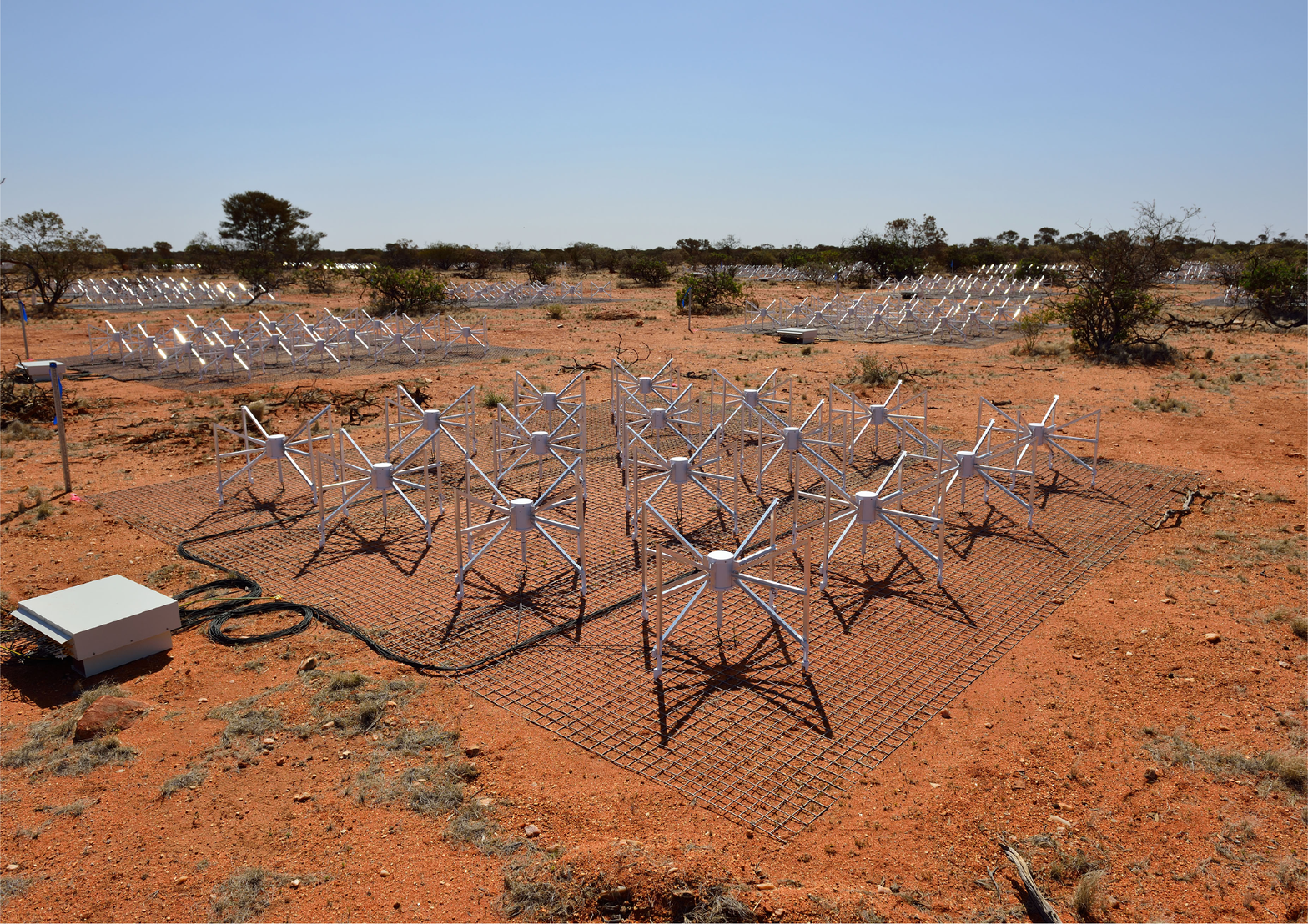}
	\caption{An MWA tile connected to a beamformer (white rectangular box). Each antenna element contains an LNA in the central hub (white cylindrical container). Photo credits: Curtin University and MWA Collaboration.\label{fig:MWA_tile}}
\end{figure}

\begin{figure}[]
	\centering
	\includegraphics[width=\linewidth]{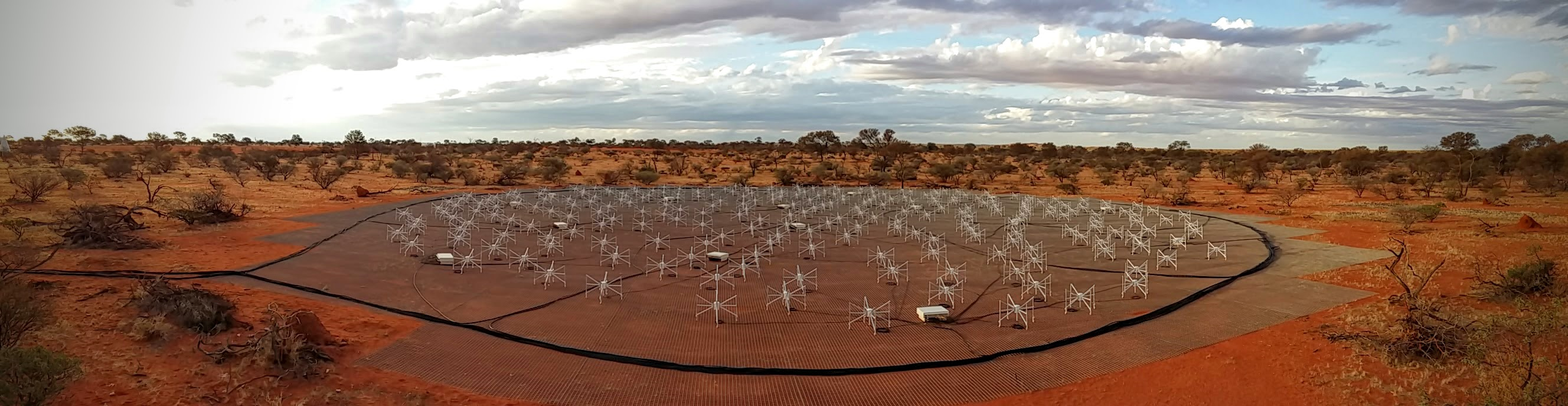}
	\caption{The Engineering Development Array. Photo credits: Curtin University and MWA Collaboration.\label{fig:EDA}}
\end{figure}

The computation of receiver noise temperature ($T_{\mathrm{rcv}}$) and radiation efficiency ($\eta_{\mathrm{rad}}$) of a phased array can be complex in nature due to the presence of mutual coupling. However, there are many frameworks for computing it. For example, the computation of $T_{\mathrm{rcv}}$ using a voltage framework was demonstrated in \cite{5062509}, while other power based methods are covered in \cite{4663504} and \cite{7079488} which involves the calculation of available gain of the phased array. For radiation efficiency calculation, power based frameworks were demonstrated in \cite{5439763,2f2cec8a372d4d888dd5a2f181e193e3 ,warnick_maaskant_ivashina_davidson_jeffs_2018}. 

The two main motivations for being able to calculate $T_{\mathrm{rcv}}$ prior to building the telescope are, the cost effectiveness of characterizing the telescope and insight into the noise coupling mechanism. Both of these abilities are useful when designing and characterizing next generation radio telescopes such as the Square Kilometre Array (SKA) \cite{5136190}. 

\added{The first contribution of this paper is to calculate the $T_{\mathrm{rcv}}$ of the Murchison Widefield Array (MWA) \cite{2013PASA...30....7T,2018PASA...35...33W} and the Engineering Development Array (EDA) \cite{2017PASA...34...34W}. The calculated $T_{\mathrm{rcv}}$ will be validated against measured results obtained via astronomical observations using similar methods seen in \cite{2007AJ....133.1505B, 2017PASA...34...34W, 7293140}. The measured $T_{\mathrm{rcv}}$ using astronomical observations also includes the effects of mutual coupling. Such comparison are not found elsewhere in literature, which forms the major contribution of this paper. 
}

The MWA is the first operational precursor telescope to the SKA-Low located at the Murchison Radio-astronomy Observatory (MRO) in the Shire of Murchison, Western Australia. The telescope consists of 256 phased arrays \cite{2018PASA...35...33W} called `tiles' as shown in Fig \ref{fig:MWA_tile}. Each tile contains 16 antenna elements, called MWA dipoles, placed in a $4 \times 4$ configuration spaced 1.1 \si{\meter} apart over a $5 \times 5$~\si{\meter} metallic ground mesh. Each element houses a low-noise amplifier (LNA) in the central hub and the output signal travels through a phase matched coaxial cable to the beamformer. The maximum spread of the tiles that make up the overall telescope is approximately 5.3 \si{\kilo\meter}. \added{The computation of $T_{\mathrm{rcv}}$ at the tile level is of interest as it} is required for determining the sensitivity of the MWA tile \cite{2013PASA...30....7T}. 

The EDA uses the same dipole elements as the MWA but houses a modified LNA\footnote{The LNA was slightly modified to have a larger bandwidth (50-300 \si{\mega\hertz}). Apart from this slight change, the LNA is identical to that used in the MWA (70-300 \si{\mega\hertz}).}. The EDA consists of 256 elements placed in a pseudo random configuration spanning 35 \si{\meter} diameter over a metallic ground mesh \cite{2017PASA...34...34W} as shown in Fig \ref{fig:EDA}. It was designed to mimic the proposed SKA-Low station configuration and therefore making it a perfect test bed as it shares nearly identical elements to the MWA. The comparison of EDA's $T_{\mathrm{rcv}}$ to MWA gives insight of the impact of mutual coupling on $T_{\mathrm{rcv}}$ which will be presented later.

\added{The $T_{\mathrm{rcv}}$ can also be measured using the Y-factor method as seen in \cite{10.1017.pasa.2014.11,7447502}, which uses a sufficiently large absorber as the hot source which covers the main beam and the sky as the cold source. At around the $\approx50-100\si{\mega\hertz}$ region, the opposite takes place as the average sky temperature is in the thousands of Kelvin. The sky is the hot source while the absorber is the cold source. However, the sky temperature exponentially decays with increasing frequency with a transition frequency at $\approx150~\si{\mega\hertz}$. At the transition frequency, $T_{hot} = T_{cold}$ and thus the Y-factor method will fail. In addition, large absorber structure is required to be built which limits the practicality of the Y-factor method to smaller arrays making unsuitable for the MWA tile and the EDA.}

\added{The second contribution of this paper is to demonstrate that a power wave based framework (PWF) found in \cite{954781} can be recast to compute $T_{\mathrm{rcv}}$ of a phased array. The fundamental concept of this framework is the computation of incident and reflected power. \added{This method was selected} because it is based in the $S$-parameters domain which is more closely connected with the authors previous work in this area \cite{8408814}.   

Additionally, the formulation can be easily modified to compute various quantities such as active/embedded reflection coefficient, transducer/available gain, and incident/reflected power of given source(s) that includes all coupling paths. An example of this will be shown by re-using this framework to calculate delivered power to the array for radiation efficiency calculation in subsequent section.}

\added{The $T_{\mathrm{rcv}}$ of an array can be calculated using \cite{5439763}}
\begin{align}
\label{eqn:T}
T_{\mathrm{rcv}} &= \frac{P_{\mathrm{av}}^{\mathrm{rcv}}}{P_{\mathrm{av}}^{\mathrm{amb}}}T_0\\
&= \frac{P_{\mathrm{av}}^{\mathrm{rcv}}}{kG_A}
\end{align}
where $P_{\mathrm{av}}^{\mathrm{rcv}}$ is the available receiver noise power at the output, $P_{\mathrm{av}}^{\mathrm{amb}}$ is the available ambient temperature noise power at the output \added{due to isotropic sky at $T_0$} \added{and $G_A$ is the available gain of the LNA for a single element but for an array, it represents the available receiver gain.}


\added{Effectively, both the previously mentioned voltage and power framework works by computing a similar ratio described by \eqref{eqn:T}. The proposed framework for calculating $T_{\mathrm{rcv}}$ using \cite{954781} was shown to be consistent with current methods in \cite{8739942} when compared to the voltage and power framework found in \cite{5062509,7079488}.} 

\added{The active reflection coefficient ($\Gamma_{\mathrm{actv}}$) alongside the input referred single element formulation given by \eqref{eqn:sing_elem} can be used as an alternative.}
\begin{align}
\label{eqn:sing_elem}
T &= T_{\mathrm{min}} + 4NT_0 \frac{|\Gamma_{\mathrm{actv}} - \Gamma_{\mathrm{opt}}|^2}{(1-|\Gamma_{\mathrm{actv}}|^2)(1-|\Gamma_{\mathrm{opt}}|^2)}
\end{align}
where the four noise parameters are represented by $T_{\mathrm{min}}$, $N$ and $\Gamma_{\mathrm{opt}}$.	
	
\added{However, the simple insight given by \eqref{eqn:sing_elem} breaks down when $|\Gamma_{\mathrm{actv}}|>1$. This over unity condition was achieved by the MWA and EDA at several pointing directions in the $50-60~\si{\mega\hertz}$ region due to the embedded reflection coefficient of the dipoles being close to unity and thus, the active reflection concept will not be discussed further in this paper aside from its links to the proposed framework found in Sect. \ref{subsec:actv_emb}.}

The remainder of this paper is organized as follows. Sect. \ref{sec:noise_calc} introduces the power wave based framework for computing receiver noise temperature and radiation efficiency, followed by results presented in Sect. \ref{sec:results}. Finally, concluding remarks are presented in Sect. \ref{sec:conlc}.

\section{Receiver and External Noise Calculation}
\label{sec:noise_calc}
\begin{figure}[]
	\centering
	\includegraphics[width=\linewidth]{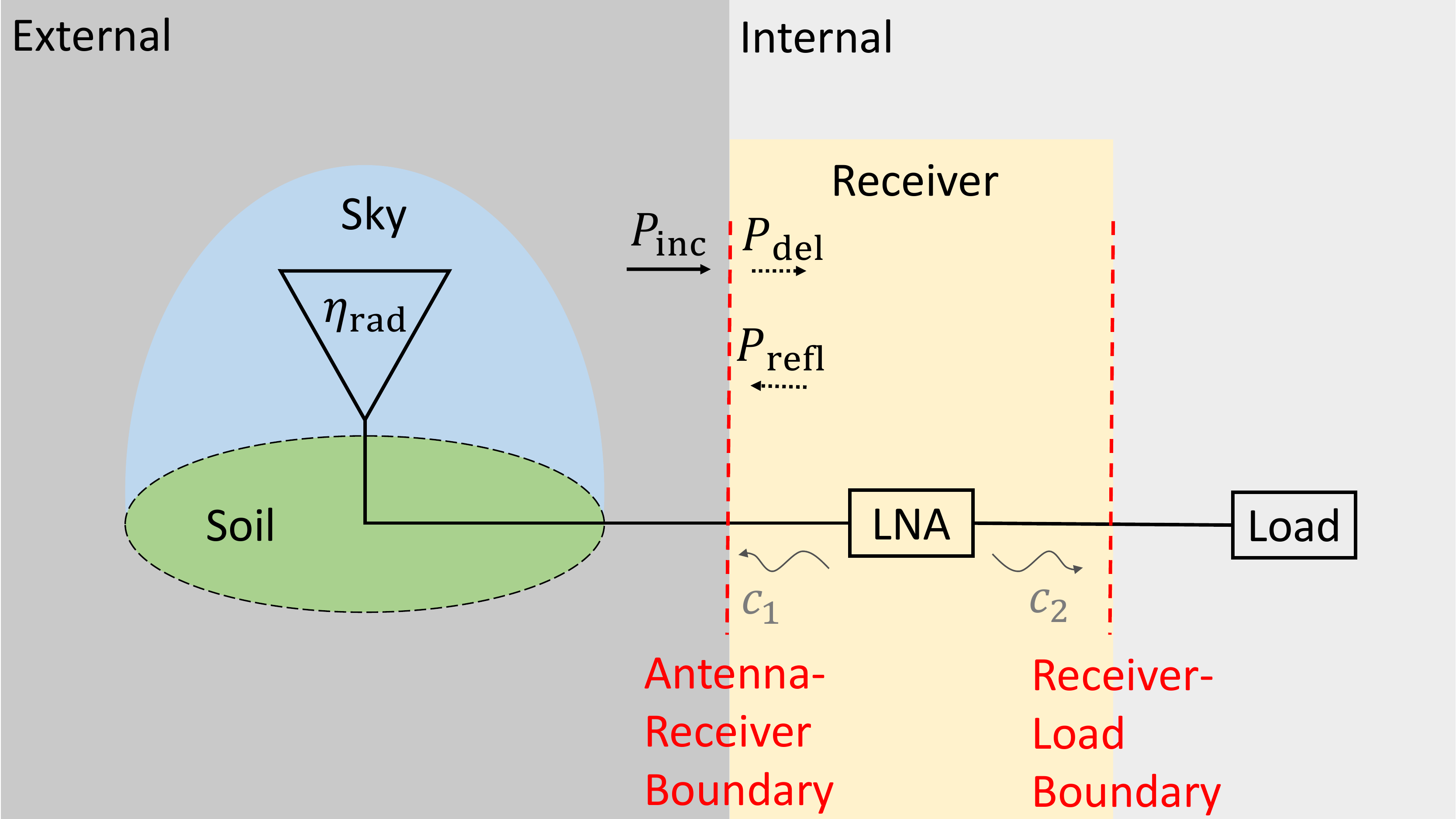}
	\caption{Overall system diagram of an antenna connected to an LNA. The system is made up of an external part which consists of an antenna, sky and soil while the internal part consists of the receiver (LNA) and a load. The boundary indicates a region in which a mismatch of impedance could occur and hence causes incident power waves to that boundary be either partially or fully reflected. The notation $P_{\mathrm{inc}}$ indicates the incident power, $P_{\mathrm{del}}$ is the delivered power, which is the difference between the incident power and the reflected power $P_{\mathrm{refl}}$. The LNA also emits noise waves at the input and output terminals labelled $c_{1}$ and $c_{2}$ respectively. The noise temperature is related to power spectral density by the relation $P=kT$ where $k$ is Boltzmann's constant and therefore they can be used interchangeably. Noise temperatures calculated involving active devices do not correspond to a physical temperature. \label{fig:overall_system}}
\end{figure}

With the aid of Fig. \ref{fig:overall_system}, let us consider the sources of noise that exist in the system. Firstly, \added{there is the} external noise which consists of noise from the sky due to naturally radiating cosmic sources, soil and thermal noise due to ohmic losses which form a net power flow that is incident onto the Antenna-Receiver Boundary represented by $P_{\mathrm{inc}}$. Secondly, \added{there is the} internal noise due to the receiver which produces noise waves indicated by $c_{1}$ and $c_{2}$ towards both boundaries \cite{168757} and noise waves emerging from the load (not shown in diagram).

For subsequent analysis, it is implied that the properties of the boundary are as follows:
\begin{enumerate}
	\item it only exists in the absence of a conjugately matched impedance with respect to the left and right hand side of the boundary, 
	\item the larger the mismatch, the more impenetrable the boundary is to the incident power wave,
	\item it is temperature invariant. That is to say, the impedance on either side of the boundary are not affected by changes in physical temperature.
\end{enumerate}

The available internal noise power ($P_{\mathrm{av}}^{\mathrm{rcv}}$) in \eqref{eqn:T} is calculated under the condition that no external noise is present. Conceptually, \added{it implies} that the antenna and load is immersed and kept at thermal equilibrium in a 0 \si{\kelvin} isotropic environment. For convenience, \added{internal noise power delivered to a noiseless reference impedance ($Z_0$) matched load was computed rather than available power at the output.} 

Similar treatment is applied for the available external noise power ($P_{\mathrm{av}}^{\mathrm{amb}}$). The delivered external noise power under the condition that no internal noise is present was calculated. Here, \added{it was} assumed that the receiver and load is immersed and kept at 0 \si{\kelvin} while the antennas are kept at thermal equilibrium in an isotropic environment at $T_0$. \added{Using the relation from \eqref{eqn:T},} it can be shown that
\begin{align}
\label{eqn:noise_temp}
T_{\mathrm{rcv}} &= \frac{P_{\mathrm{int}}^{\mathrm{out}}}{kG_{T}}\\
\label{eqn:GT}
G_{T} &= \frac{P_{\mathrm{ext}}^{\mathrm{out}}}{kT_0}
\end{align}  
where $P_{\mathrm{int}}^{\mathrm{out}}$ is the noise power delivered to a noiseless matched load due to internal sources alone, $P_{\mathrm{ext}}^{\mathrm{out}}$ is the noise power delivered to a noiseless matched load due to external sources alone and $G_{T}$ is the \added{receiver transducer gain}. The transducer gain is defined as the ratio of the delivered power by the network to the available power from the source ($kT_0$) \cite{gonzalez1997microwave}.

This analysis is simple for a single isolated element as the quantity $P_{\mathrm{int}}^{\mathrm{out}}$ and $P_{\mathrm{ext}}^{\mathrm{out}}$ are easily computed. However, for an array of closely spaced antennas this is no longer the case. In this scenario, the antennas mutually couple and causes $P_{\mathrm{int}}^{\mathrm{out}}$ to deviate away from a single element case. The mechanisms that cause this overall effect are
\begin{enumerate}
	\item changing of embedded antenna impedance, 
	\item coupling of outbound internal noise to neighbouring elements.
\end{enumerate}
Furthermore, the computation of $P_{\mathrm{ext}}^{\mathrm{out}}$ and subsequently $G_{T}$ for an array is not apparent at first glance due to the complex coupling paths. 

\subsection{Contribution of Internal Noise}
\label{subsec:Int_noise}
\begin{figure}[]
	\centering
	\includegraphics[width=\linewidth]{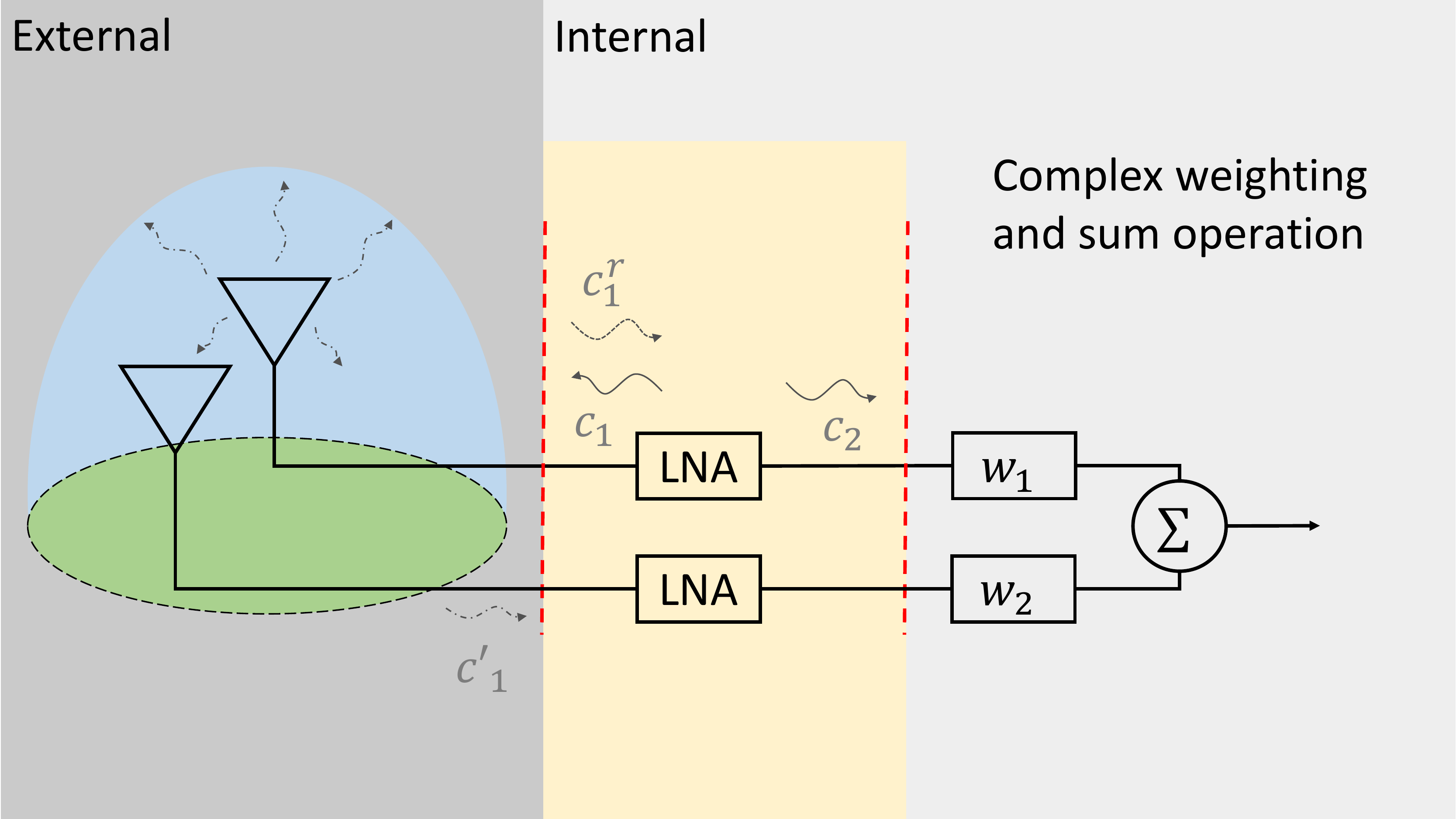}
	\caption{Coupling path of internal noise sources alone for a two-element array. The load as seen in Fig. 
		\ref{fig:overall_system} is now replaced with a complex weight and sum operator; however, the assumption of matched condition still remains. The output referred receiver noise temperature consists of reflected wave $c^r_{1}$, coupled wave to neighbouring element $c_1'$ and noise wave $c_{2}$ emanating from the output of the receiver. While not shown, similar coupling paths occur at the lower branch.  \label{fig:two-element}}
\end{figure}

Fig. \ref{fig:two-element} shows an example of noise paths that each noise wave will undergo for a two-element array. \added{All these various paths can be accounted for} using matrices to compute the outgoing receiver noise power at the boundaries as follows
\begin{align}
\label{eqn:P_int_randa}
\mathbf{A}_{\mathrm{int}}^{\mathrm{out}} &= \mathbf{M}\mathbf{\hat{N}}\mathbf{M}^\dagger\\
\label{eqn:M}
\mathbf{M} &= \left[\mathbf{I}-\mathbf{S}_{\mathrm{LNA}}\mathbf{S}_{\mathrm{load}}\right]^{-1}
\end{align}
where $\mathbf{A}_{\mathrm{int}}^{\mathrm{out}}$ is a $n\times n$ noise correlation Hermitian matrix of the outgoing noise power from each port (inputs and outputs) for an $n$-port network, $\mathbf{\hat{N}}$ is the noise correlation matrix of the multiport amplifier due to internal sources alone, $\mathbf{M}$ accounts for the mismatches in impedance, ${\cdot}^\dagger$ is the Hermitian operator, $\mathbf{S}_{\mathrm{LNA}}$ and $\mathbf{S}_{\mathrm{load}}$ are the $S$-parameters of the multiport amplifier and the combined source (antenna) and load network attached to the multiport amplifier respectively. For completeness, the derivation of $\mathbf{M}$ can be found in Appendix \ref{sec:mismatch}.
\begin{figure}[t]
	\centering
	\includegraphics[width=0.8\linewidth]{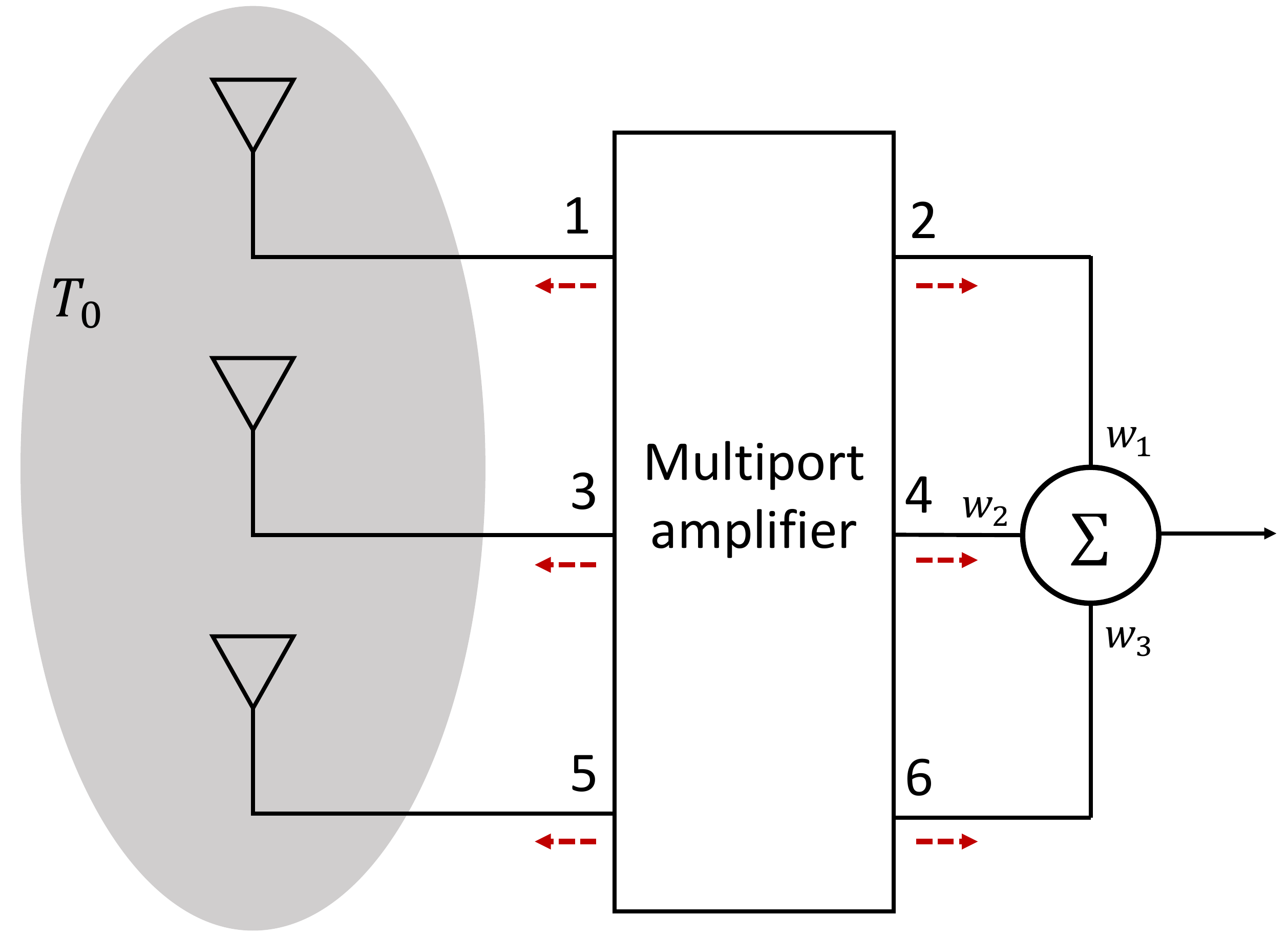}
	\caption{Example port numbering convention for a three element phased array. The odd numbered ports are the input ports of the multiport amplifier connected to antennas while the even numbered ports are the output ports. The noise waves due to internal sources alone emerging from the multiport network are described by \eqref{eqn:P_int_randa} and are represented by dashed arrows. Noise waves emerging at the output ports undergo a weight ($w_i$) and sum operation.  \label{fig:port_numbering}}
\end{figure}

The noise power wave quantities computed by \eqref{eqn:P_int_randa} can be visualized with the aid of Fig. \ref{fig:port_numbering}. Each entry of the matrix contains information of outbound noise power waves (dashed arrows) due to internal sources alone after all coupling paths have been accounted for. The main diagonal contains the total noise power emerging from the network ports due to all $c_1$ and $c_2$ sources, while the cross terms contain the total amount of correlated power between port $m$ and $n$.

To compute \eqref{eqn:P_int_randa} correctly, $\mathbf{S}_{\mathrm{LNA}}$, $\mathbf{\hat{N}}$ and $\mathbf{S}_{\mathrm{load}}$ \added{must have} a consistent port numbering convention. Based on the port numbering convention shown in Fig. \ref{fig:port_numbering}, assuming that 
\begin{enumerate}
	\item the multiport amplifier is constructed from identical isolated 2-port element LNAs\footnote{Non-identical LNAs can also be used by modifying the entries in \eqref{eqn:S_lna} and \eqref{eqn:N_hat} to include the measured or simulated parameters of the non-identical LNA multiport network.},
	\item odd numbered ports are inputs and even numbered ports are outputs of the multiport amplifier network,
	\item a reflectionless load ($Z_0$) is attached to the outputs of the LNAs then,
\end{enumerate}
\begin{align}
\label{eqn:S_lna}
\mathbf{S}_{\mathrm{LNA}} &= \left[ 
	{\begin{array}{ccccc}	
		S_{11} & S_{12} & 0 & 0 & \dots   \\ 
		S_{21} & S_{22} & 0 & 0 & \dots  \\
		 0 & 0 & S_{11} & S_{12} & \dots \\ 
		 0 & 0 & S_{21} & S_{22} & \dots\\
		 \vdots &  \vdots &  \vdots & \vdots & \ddots  \\ 
		 0 & 0 & 0 & 0 & \dots \\ 
		 0 & 0 & 0 & 0 & \dots \\
		\end{array}} \right]\\
\label{eqn:N_hat}
\mathbf{\hat{N}} &= \left[ 
	{\begin{array}{ccccc}	
		\langle|c_1|^2\rangle & \langle c_1c_2^* \rangle & 0 & 0 & \dots  \\ 
		\langle c_1^*c_2 \rangle & \langle|c_2|^2\rangle & 0 & 0 & \dots \\
		0 & 0 & \langle|c_1|^2\rangle & \langle c_1c_2^* \rangle & \dots\\ 
		0 & 0 & 	\langle c_1^*c_2 \rangle &\langle|c_2|^2\rangle & \dots\\
		\vdots &  \vdots &  \vdots & \vdots & \ddots  \\ 
		0 & 0 & 0 & 0 & \dots \\ 
		0 & 0 & 0 & 0 & \dots\\
		\end{array}} \right]\\
\label{eqn:S_load}	
\mathbf{S}_{\mathrm{load}} &= \left[ 
{\begin{array}{cccccc}	
	S_{11}^\mathrm{ant} & 0 & S_{12}^\mathrm{ant} & \dots & S_{1,n}^\mathrm{ant} & 0  \\ 
	0 & 0 & 0 & 0 & 0 & 0\\ 
	S_{21}^\mathrm{ant} & 0 & S_{22}^\mathrm{ant} & \dots & S_{2,n}^\mathrm{ant} & 0  \\
   \vdots & 0 & \vdots & \ddots & \vdots & 0  \\
	S_{m,1}^\mathrm{ant} & 0 & S_{m,2}^\mathrm{ant} & \dots & S_{mm}^\mathrm{ant} & 0\\
	\added{0} & \added{0} & \added{0} & \added{0} & \added{0} & \added{0}\\
	\end{array}} \right].
\end{align}

For the computation of $T_{\mathrm{rcv}}$ seen in \eqref{eqn:noise_temp}, the delivered noise power to loads at the output of the network \added{are contained in the subset of matrix} \eqref{eqn:P_int_randa}. As the input ports to be odd numbered, all the odd rows and columns to form a submatrix $\left[\mathbf{P}_{\mathrm{int}}^{\mathrm{out}}\right]_{m,n=2,4,\dots,m}$ can be removed. \added{This submatrix is the correlation matrix that can be used for characterizing the amount of noise coupling that exists in correlating arrays}. To get the total coupled noise power after the summer at the output of a phased array, this submatrix is multiplied by the beamformer weights as follow
\begin{equation}
\label{eqn:P_int_randa_tot}
P_{\mathrm{int}}^{\mathrm{out}}= \mathbf{w}\left[\mathbf{A}_{\mathrm{int}}^{\mathrm{out}}\right]_{m,n=2,4,\dots,m}\mathbf{w}^\dagger
\end{equation}
where $\mathbf{w}$ is a row vector containing the applied beamformer complex weights and $\{\mathbf{A}_{\mathrm{int}}^{\mathrm{out}}\}_{m,n=2,4,\dots,m}$ a submatrix containing even numbered rows and columns of \eqref{eqn:P_int_randa}. To ensure correct scaling in the calculated output power, the amplitude of $\mathbf{w}$ is scaled by the number of elements $N$ such that $\sum_{i=1}^{N}|\mathbf{w}_{i}|^2 = 1$.

While the port convention in \eqref{eqn:S_lna}-\eqref{eqn:S_load} is not unique, this convention was chosen \added{as it is easier to construct/modify the required matrices}. \added{The focus now shifts} to the computation of transducer gain for a phased array. 

\subsection{Contribution of External Noise}
\label{subsec:Ext_noise}
The outgoing power wave at the inputs and outputs of the multiport network due to power incident at the input ports are given by  
\begin{align}
\label{eqn:P_ext_randa}
\mathbf{A}_{\mathrm{ext}}^{\mathrm{out}} =\mathbf{M}\mathbf{S}_{\mathrm{LNA}}\mathbf{\hat{aa}}^\dagger(\mathbf{S}_{\mathrm{LNA}})^\dagger\mathbf{M}^\dagger
\end{align}
where $\mathbf{\hat{aa}}^\dagger$  contains the noise correlation matrix of the attached loads at the input and output of the LNA.

From the perspective of a matched load at the output ports of the multiport network, \eqref{eqn:P_ext_randa} describes the incident power from the network at the Receiver-Load Boundary. On the other hand, from the perspective of the source at the input ports, \eqref{eqn:P_ext_randa} describes the reflected power at the Antenna-Receiver Boundary (see Fig. \ref{fig:port_numbering}).

For passive loads at thermal equilibrium with $T_0$ the noise correlation matrix $\mathbf{\hat{aa}}^\dagger$ can be determined by Bosma's theorem \cite{89082} which states that
\begin{equation}
\label{eqn:Bosma}
\mathbf{\hat{aa}}^\dagger = kT_0\left[\mathbf{I}-\mathbf{S}_{\mathrm{load}}(\mathbf{S}_{\mathrm{load}})^\dagger\right].
\end{equation}

To simulate noiseless loads being attached at the output of the LNA, $\mathbf{I}$ in \eqref{eqn:Bosma} must be replaced by
\begin{align}
\label{eqn:Bosma_mod}
\mathbf{I'} =\resizebox{0.4\linewidth}{!}{$\left[ 
	{\begin{array}{cccccc}	
		1 & 0 & 0 & \dots & 0 & 0  \\ 
		0 & 0 & 0 & \dots & 0 & 0  \\ 
		0 & 0 & 1 & \dots & 0 & 0  \\ 
		\vdots & 0 & 0 & \ddots & 0 & \vdots  \\ 
		0 & 0 & 0 & \dots & 1 & 0  \\ 
		0 & 0 & 0 & \dots & 0 & 0  \\ 
		\end{array}} \right].$}
\end{align}
The assumption of noiseless loads being attached to the output stage does not introduce a measurable change as a well-designed receiver chain should be dominated by LNA noise. The total delivered power from the network to $Z_0$ load is given by
\begin{equation}
\label{eqn:P_ext_randa_tot}
P_{\mathrm{ext}}^{\mathrm{out}}= \mathbf{w}\left[\mathbf{A}_{\mathrm{ext}}^{\mathrm{out}}\right]_{m,n=2,4,\dots,m}\mathbf{w}^\dagger
\end{equation}

While not shown, \eqref{eqn:P_int_randa_tot} and \eqref{eqn:GT} produces identical results to formulations found in \cite{7079488} for the computation of incident noise power and transducer gain\footnote{In \cite{7079488}, the transducer gain is called effective available gain.}.

\subsection{\added{Links to Active Reflection Coefficient Concept}}
\label{subsec:actv_emb}
\added{Before proceeding further, it is worth discussing links to theory presented in Sect. \ref{subsec:Int_noise} and \ref{subsec:Ext_noise} to the commonly used active reflection formulation. 

The output referred noise temperature calculated using the active reflection coefficient \eqref{eqn:sing_elem} relates to \eqref{eqn:P_int_randa_tot} via
\begin{align}
\label{eqn:p_ext_relate}
\mathbf{w}\left[\mathbf{A}_{\mathrm{int}}^{\mathrm{out}}\right]_{m,n=2,4,\dots,m}\mathbf{w}^\dagger & = k\frac{1}{M} \sum^{M}_{i=1} T_{i}G_{T,i}\\
\label{eqn:GT_actv}
G_{T,i} &= \frac{1-|\Gamma_{\mathrm{actv},i}|^2}{|1-S_{11}\Gamma_{\mathrm{actv},i}|^2}|S_{21}|^2
\end{align}
where $M$ is the number of elements in the array, $T_{i}$ and $G_{T,i}$ are the input referred noise temperature and the transducer gain of the $i^{th}$ element calculated using \eqref{eqn:sing_elem} and \eqref{eqn:GT_actv} respectively.

In addition, \eqref{eqn:P_ext_randa_tot} can be calculated from \eqref{eqn:GT_actv} using
\begin{align}
\label{eqn:GT_actv_relate}
 \mathbf{w}\left[\mathbf{A}_{\mathrm{ext}}^{\mathrm{out}}\right]_{m,n=2,4,\dots,m}\mathbf{w}^\dagger = \frac{1}{M} \sum^{M}_{i=1} G_{T,i}.
\end{align}

Substituting \eqref{eqn:p_ext_relate} and \eqref{eqn:GT_actv_relate} into \eqref{eqn:noise_temp} and simplifying yields
\begin{align}
\label{eqn:Trcv_actv}
T_{\mathrm{rcv}}= \frac{\sum^{M}_{i=1} T_{i}G_{T,i}}{\sum^{M}_{i=1} G_{T,i}}.
\end{align}

The $T_{\mathrm{rcv}}$ calculated using \eqref{eqn:Trcv_actv} is exact. This calculation becomes an approximation when attempting to solely use \eqref{eqn:sing_elem} to infer the array noise temperature for cases when $G_{T,m} \neq G_{T,n} \cdots \neq G_{T,z}$ and/or $|\Gamma_{\mathrm{actv}}|$ is greater than unity. Under this condition the average input referred noise temperature diverges away from the array noise temperature and therefore, it is more general to discuss the behaviour of the array's output referred noise temperature and the transducer gain separately.}

\subsection{Radiation Efficiency Calculation}
\label{subsec:ext_noise}
Radiation efficiency ($\eta_{\mathrm{rad}}$) of any antenna structure is defined by \cite{5439763,2f2cec8a372d4d888dd5a2f181e193e3}
\begin{align}
\label{eqn:rad_effiency}
\eta_{\mathrm{rad}} &= \frac{P_{\mathrm{rad}}}{P_{\mathrm{inj}}}\\
P_{\mathrm{rad}}  &=  \frac{1}{2Z_{\eta_{0}}}\int_{0}^{2\pi}\int_{0}^{\pi}\mathbf{E}_{\mathrm{ff}}\cdot\mathbf{E}_{\mathrm{ff}}^\dagger\sin \theta d\theta d\phi\\
P_{\mathrm{inj}}  &= \frac{1}{2}\mathbb{R}\{V_{\mathrm{ant}}I^*\}
\end{align} 
where $P_{\mathrm{rad}}$ is the total radiated power and $P_{\mathrm{inj}}$ is the total injected power into the antenna, $Z_{\eta_{0}}$ is the free space impedance and $\mathbf{E}_{\mathrm{ff}}$ is the far-field embedded element pattern (EEP) of the antenna as a function of $\theta$ and $\phi$, $V_{\mathrm{ant}}$ is the voltage drop across the antenna, $I$ is port current, $\{\cdot\}$ and $\{^*\}$ represents the element by element multiplication and complex conjugate operator respectively.

As noted in \cite{2f2cec8a372d4d888dd5a2f181e193e3}, the accuracy of this computation for high efficiency arrays is limited by the numerical sampling and integration of the far-field pattern used to compute $P_{\mathrm{rad}}$. Furthermore, additional data such as port currents and voltage drop across the antenna terminals are required to be saved for the computation of $P_{\mathrm{inj}}$. While the numerical integration is unavoidable, the aim is to reduce the amount of additional data required to be saved and reuse pre-existing data obtained during the characterization of the phased array such as embedded element pattern and $S$-parameter simulation. This provides the additional motivation for this section. 

\begin{figure}[t]
	\centering
	\includegraphics[height= 5cm]{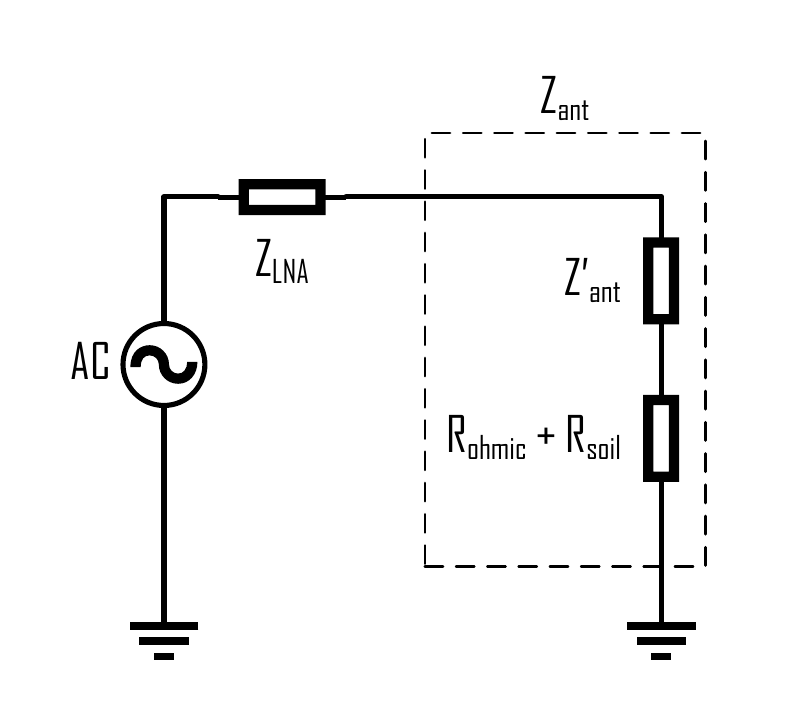}
	\caption{Simplified equivalent circuit of a lossy antenna loaded with LNA impedance in transmit mode. The $Z_{\mathrm{ant}}$ is the antenna impedance as obtained by $S$-parameter simulation/measurement or can be obtained via the port currents from simulation as both the excitation voltage and LNA impedance are known. The ohmic and soil losses are modelled as a resistor but in practice, these losses have a more complex form. \label{fig:equiv_circ}}
\end{figure}
A different approach to compute $\eta_{\mathrm{rad}}$ is to use the pattern overlap integral (POI) method found in \cite{5439763}, which eliminates the need to know the injected power by reversing the problem from a transmit to receive antenna. The EEPs required for the POI method are based on open circuit condition of all neighbouring elements whereas, EEPs generated in \cite{7731419} are based on loaded condition (LNA input impedance) of all elements. To reuse pre-existing EEPs generated in \cite{7731419}, POI formulation \added{requires modification}. 

Fig. \ref{fig:equiv_circ} shows an equivalent circuit of a transmit antenna. The total radiated power captured in the far-field pattern is due to the power dissipated by $Z'_{\mathrm{ant}}$. This means that the far-field EEP does not directly contain the knowledge of any losses. By reversing the problem to receive mode, the noise power delivered to $Z_{\mathrm{LNA}}$ by $Z'_{\mathrm{ant}}$ at a nominal physical temperature $T_0$ can be determined for a given EEP generated in transmit mode. The total noise power delivered to the \added{array terminated with identical LNA impedances} given that the antenna sees an homogeneous sky at $T_0$ is given by 
\begin{align}
\label{eqn:pdel_dash_lna}
P'_{\mathrm{LNA}} &= kT_0\mathbf{w}\mathbf{L}^{sky}\mathbf{w}^\dagger\\
L^{sky}_{m,n} &= \mathbb{R}\{\frac{Z_{\eta_{0}}}{Z_{\mathrm{LNA}}\lambda^2}\}\int_{0}^{2\pi}\int_{0}^{\pi}\mathbf{l}_{\mathrm{p},m}\cdot\mathbf{l}_{\mathrm{p},n}^\dagger\sin \theta d\theta d\phi 
\end{align} 
where $\mathbf{L}^{sky}$ is the noise correlation matrix due to homogeneous sky that is based on the POI formulation, $\lambda$ is the wavelength in meters, and $\mathbf{l}_{\mathrm{p},i}$ relates the incident electromagnetic wave to the voltage seen at the load of the $i^{\mathrm{th}}$ element as a function of $\theta$ and $\phi$. The derivation of $\mathbf{l}_{\mathrm{p}}$ can be found in Appendix \ref{sec:norm_pat}.

The total noise power delivered to $Z_{\mathrm{LNA}}$ by $Z_{\mathrm{ant}}$ at physical temperature of $T_0$ \added{can be calculated} using Bosma's theorem \cite{89082} or Twiss's theorem \cite{5439763}. For our purposes, Bosma's theorem is a more convenient choice as it uses $S$-parameter natively. This is where the versatility of \cite{954781} comes into play. The formulation \added{can easily be modified} to calculate the total delivered external noise power at the input of the LNAs using
\begin{align}
\label{eqn:pdel_lna}
P_{\mathrm{LNA}} &= \mathbf{w}\{\mathbf{A}_{\mathrm{ext}}^{\mathrm{inc}}-\mathbf{A}_{\mathrm{ext}}^{\mathrm{out}}\}_{m,n=1,3,\dots,m}\mathbf{w}^\dagger\\
\label{eqn:p_inc_ext}
\mathbf{A}_{\mathrm{ext}}^{\mathrm{inc}} &=  \mathbf{M}'\mathbf{\hat{aa}}^\dagger\mathbf{M}'^\dagger\\
\label{eqn:M_dash}
\mathbf{M}' &= \left[\mathbf{I}-\mathbf{S}_{\mathrm{load}}\mathbf{S}_{\mathrm{LNA}}\right]^{-1}
\end{align} 
where $\mathbf{A}_{\mathrm{ext}}^{\mathrm{inc}}$ is the incident noise power to the LNA due to external sources alone, $\mathbf{A}_{\mathrm{ext}}^{\mathrm{out}}$ and $\mathbf{\hat{aa}}^\dagger$ were previously computed in \eqref{eqn:P_ext_randa} and \eqref{eqn:Bosma}.

The radiation efficiency is then given by 
\begin{align}
\label{eqn:rad_eff}
\eta_{\mathrm{rad}} = \frac{P'_{\mathrm{LNA}}}{P_{\mathrm{LNA}}}
\end{align} 

This is effectively how the POI method works. For verification, it was shown in \cite{8888935} that the radiation efficiency calculated using \eqref{eqn:rad_eff} and \eqref{eqn:rad_effiency} produced identical results within numerical error. The result is reproduced here with additional verification using perfect electric conductor (PEC) materials.
\begin{figure}[t]
	\centering
	\includegraphics[width=\linewidth]{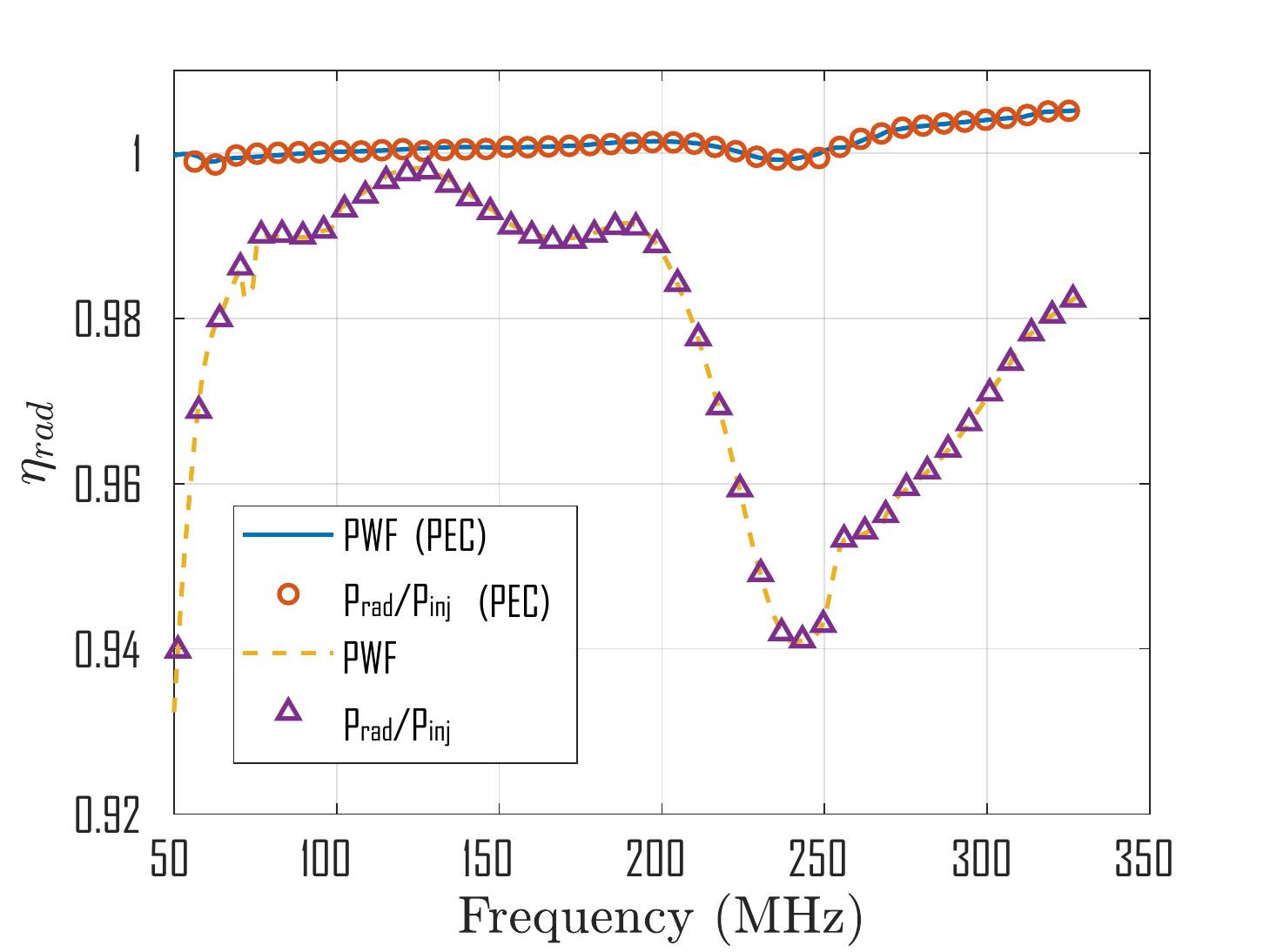}
	\caption{Calculated efficiency of the MWA at zenith. The PWF and $P_{\mathrm{rad}}/P_{\mathrm{inj}}$ curve represents the efficiency calculated using the power wave framework and \eqref{eqn:rad_effiency} respectively. The EEPs used to compute both results are based on prior simulation presented in \cite{7731419}, whereby the radiation efficiency is solely due to soil losses as all the metallic elements used during simulation were perfect electric conductors (PEC). For further validation, the MWA was re-simulated with PEC infinite ground to ensure that $\eta_{\mathrm{rad}}=1$ is obtained with the PWF and \eqref{eqn:rad_effiency}. \label{fig:MWA_efficiency}}
\end{figure}

The efficiency of the MWA as seen in Fig. \ref{fig:MWA_efficiency} was calculated using \eqref{eqn:rad_eff}. For result verification, these values were compared against those obtained using \eqref{eqn:rad_effiency}. The numerical integration on the radiation pattern was performed at a resolution of $0.2^\circ$ for both methods. Additionally, previous raw simulation data was reprocessed to obtain the port currents required for the computation of $P_{\mathrm{inj}}$. 

For the PEC case, the efficiency reported was as high as 100.5\% which is due to the limitations of the simulation package. As demonstrated in \cite{2f2cec8a372d4d888dd5a2f181e193e3}, the efficiency obtained using Method-of-Moments (MoM) based solver such as FEKO for PEC materials ranges from 100\% to 100.7\%. Efficiency calculation was not done for the EDA as the array was simulated using perfect electric conductor (PEC) material over infinite ground and due to lengthy simulation time, it was not repeated with lossy materials.  

\section{Results}
\label{sec:results}
\subsection{Receiver Temperature}
\begin{figure}[]
	\centering
	\includegraphics[width=\linewidth]{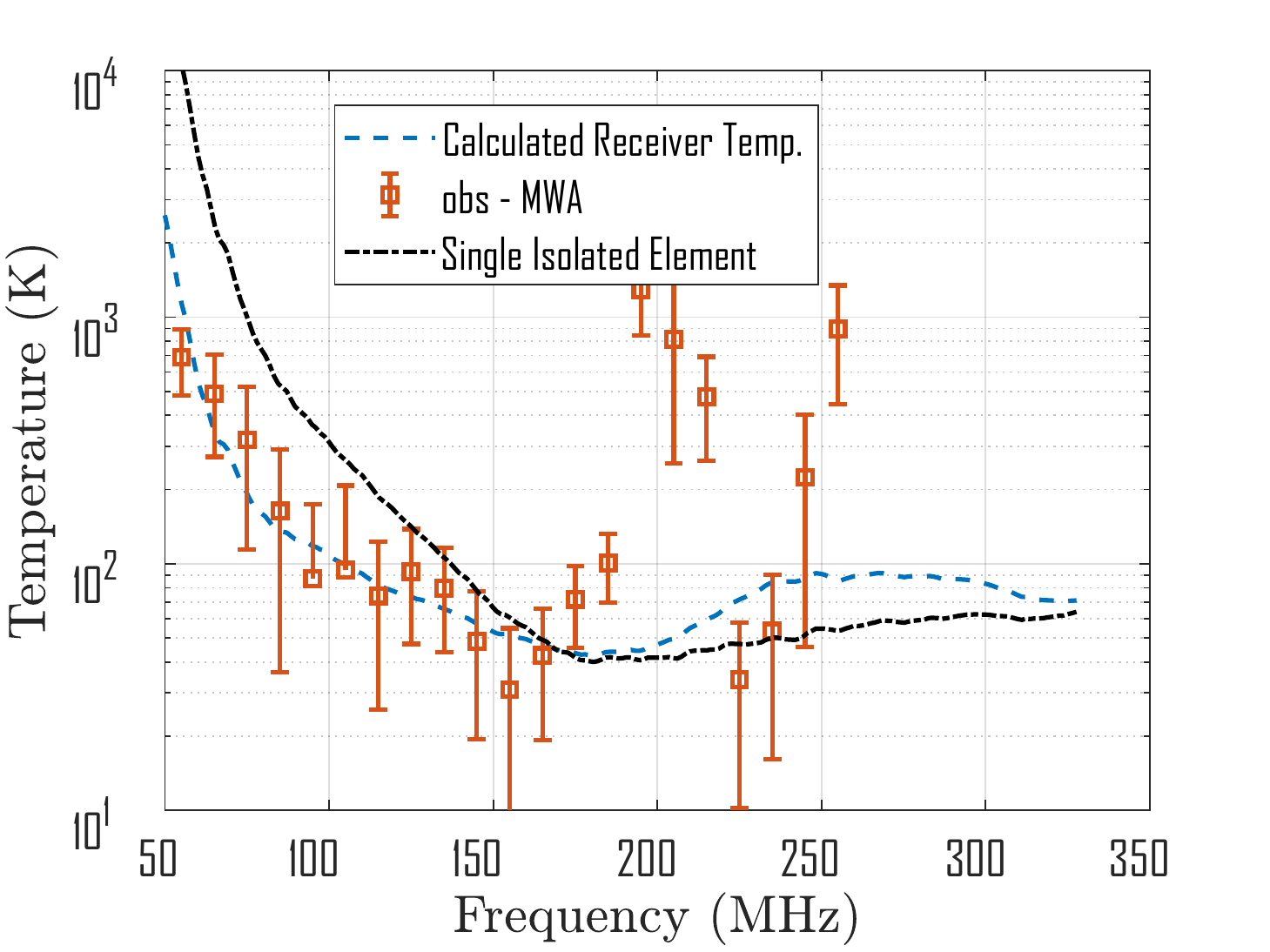}
	\caption{Comparison between the calculated (dashed curve) and observed receiver noise temperature of MWA. The mean and standard deviation of the observed receiver noise temperature represented by the data points was calculated over 128 tiles. A single isolated element result represented by the dot-dash curve is presented for comparison.\label{fig:MWA_comparison}}
\end{figure}

\begin{figure}[]
	\centering
	\includegraphics[width=\linewidth]{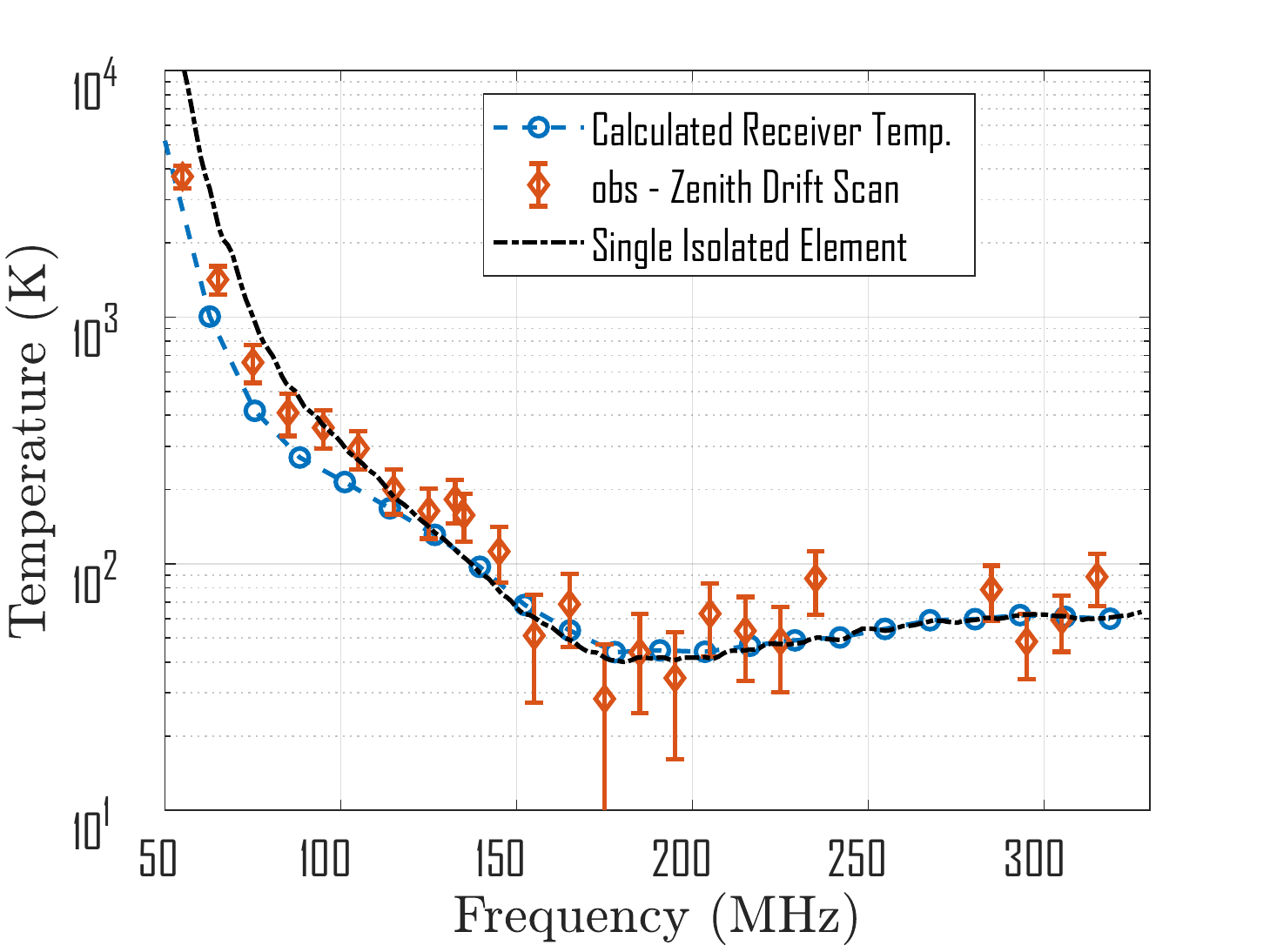}
	\caption{Comparison between the calculated and observed receiver noise temperature of the EDA. The calculated result represented by the dashed curve is based on a fully perfect electric conductor (PEC) array and therefore the efficiency calculation was not performed nor included. A single isolated element result represented by the dot-dash curve is also presented for comparison. The observed receiver noise temperature represented by the data points shows the mean and standard deviation of the receiver noise temperature taken over a 1 \si{\MHz} bin size. \label{fig:EDA_comparison}}
\end{figure}

The MWA tile and the EDA were simulated using an electromagnetic simulator FEKO to obtain the $S$-parameter of the arrays. \added{The expected $T_{\mathrm{rcv}}$ and $\eta_{\mathrm{rad}}$ was calculated for the two arrays using \eqref{eqn:t_rcv_obs} and the measured noise and $S$-parameters of the LNA obtained in \cite{8408814}.} The calculated receiver noise temperature was then compared to values obtained via astronomical drift scan method as seen in \cite{2007AJ....133.1505B, 2017PASA...34...34W, 7293140}. 

Fig. \ref{fig:MWA_comparison} shows the comparison between the calculated and observed receiver noise temperature of the MWA. The observations were performed on the 9th June 2014 by setting all the MWA beamformes to point overhead (at zenith) and allowing astronomical sources to drift through the MWA's beam. The expected power detected as a function of frequency ($\nu$) by a phased array is given by
\begin{align}
\label{eqn:power_obs}
P(\nu) &= g(\nu)k\left[\eta_{\mathrm{rad}}T_{\mathrm{ant}}(\nu) + T'_{\mathrm{rcv}}(\nu)\right]\\
\label{eqn:t_rcv_obs}
T'_{\mathrm{rcv}}(\nu) &= T_{\mathrm{rcv}}(\nu) + \left[1-\eta_{\mathrm{rad}}(\nu)\right]T_a
\end{align}  
where $g(\nu)$ is the \added{overall power gain of the array signal chain which includes the transducer gain, cable losses, secondary amplication stage etc}, $T_{\mathrm{ant}}$ is the antenna temperature due to sky noise and $T_a$ is the ambient temperature.

\added{The quantity $T'_{\mathrm{rcv}}(\nu)$ can be obtained by first} modelling the predicted power as
\begin{align}
\label{eqn:power_model}
P'(\nu) &= g(\nu)k\left[\eta_{\mathrm{rad}}T_{\mathrm{ant}}^{\mathrm{model}}(\nu) + T'_{\mathrm{rcv}}(\nu)\right]
\end{align}
and $T_{\mathrm{ant}}^{\mathrm{model}}$ is estimated as
\begin{align}
\label{eqn:sky_integration}
T_{\mathrm{ant}}^{\mathrm{model}}(\nu) &= \frac{\int_{\Omega} B(\nu,\theta,\phi) T(\nu,\theta,\phi) d\Omega}{\int_{\Omega} B(\nu,\theta,\phi) d\Omega}
\end{align}
where $B(\nu,\theta,\phi) = E(\nu,\theta,\phi)_{\mathrm{ff}}\cdot E(\nu,\theta,\phi)_{\mathrm{ff}}^\dagger$ is the simulated far-field power pattern as a function of $\theta$ and $\phi$, $T(\nu,\theta,\phi)$ is the sky brightness temperature obtained from "Haslam Map" \cite{1982A&AS...47....1H} at frequency $\nu$, which has been scaled down from the original $408$~\si{\MHz} to lower frequencies by multiplying by a factor $(\nu/408~\si{\mega\hertz})^{-2.55}$.

Least square optimization is then performed on the predicted $P'(\nu)$ with the observed $P(\nu)$ to solve for $g(\nu)$ and $T'_{\mathrm{rcv}}(\nu)$ respectively. Based on \eqref{eqn:power_model} the power received by every tile is expected to be proportional to $T_{\mathrm{ant}}^{\mathrm{model}} + T'_{\mathrm{rcv}}(\nu)$, assuming that the sky model used in \eqref{eqn:sky_integration} is a good representation of the true sky (accuracy of sky model is of order $\approx 10 \%$). This relation has been identified to hold best for the 12 to 14 hours range of the Local Sidereal Time (LST) \footnote{LST is an hour angle between vernal equinox and local meridian} (see also \cite{2017PASA...34...34W} for a more detailed justification of LST range selection procedure). Once the calculated values had been obtained, this model is fitted to measured data using least squares to solve for $g(\nu)$ and $T'_{\mathrm{rcv}}(\nu)$.

The frequency ranges of $170$~\si{\MHz} to $220$ \si{\MHz} and $240$ \si{\MHz} and above show radio frequency interference (RFI) which causes the observed $T'_{\mathrm{rcv}}$ to increase dramatically. The error bars generated are based on the standard deviation of $T'_{\mathrm{rcv}}$ calculated over $128$ MWA tiles used during observation.

Fig \ref{fig:EDA_comparison} shows the comparison between the calculated and observed  $T'_{\mathrm{rcv}}(\nu)$ of the EDA. Detailed $T'_{\mathrm{rcv}}$ calculation from astronomical observation can be found in \cite{2017PASA...34...34W}.

\subsection{Transducer Gain}
\begin{figure}[]
	\centering
	\includegraphics[width=\linewidth]{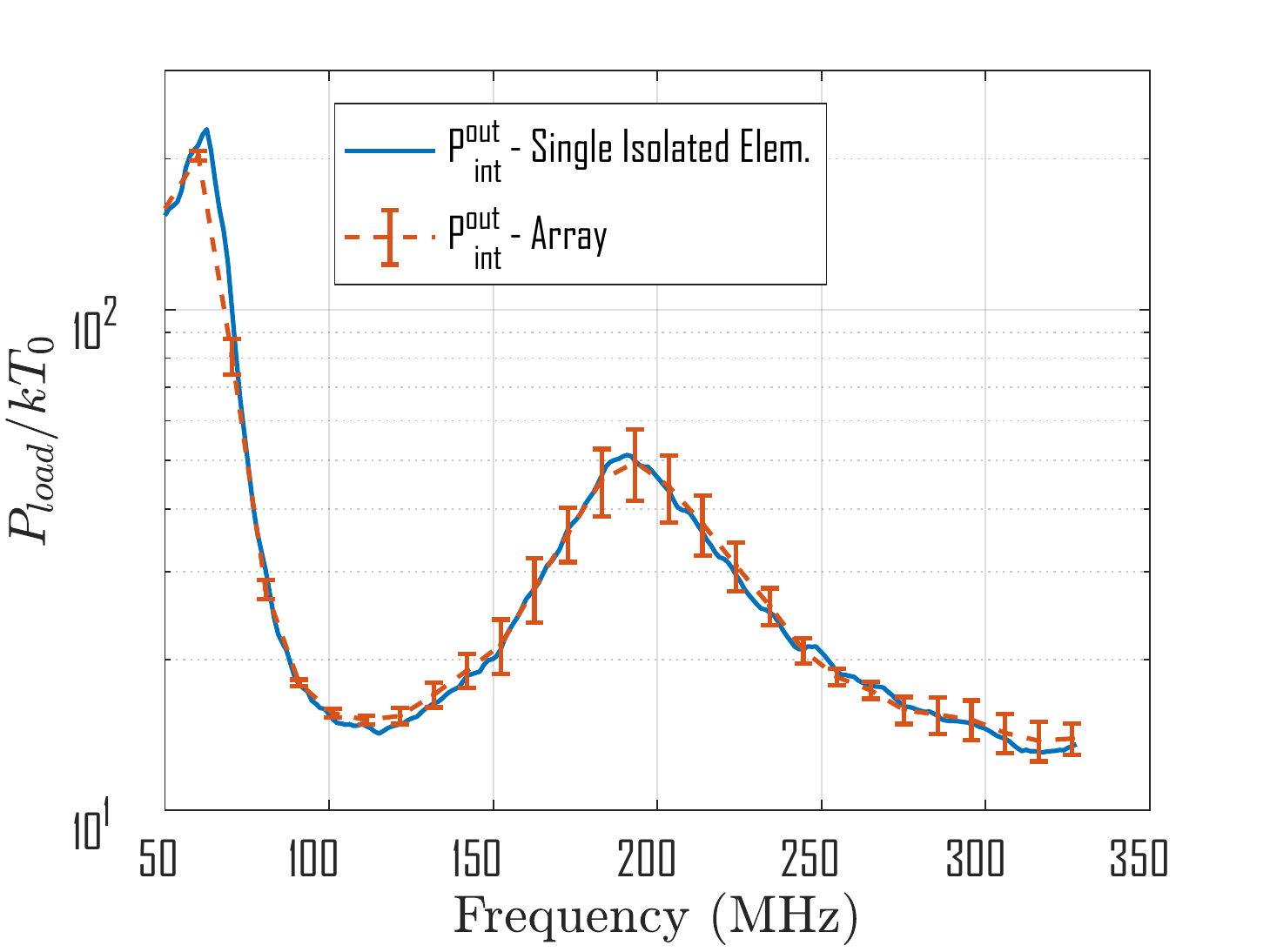}
	\caption{Delivered noise power to a $Z_0$ matched load due to internal sources alone normalized to $kT_0$. The solid line represents the delivered noise power by a single isolated MWA element whereas the dashed line represents the mean and standard deviation of power delivered by the MWA array over 197 optimal MWA pointings.\label{fig:pint_comparison}}
\end{figure}

\begin{figure}[]
	\centering
	\includegraphics[width=\linewidth]{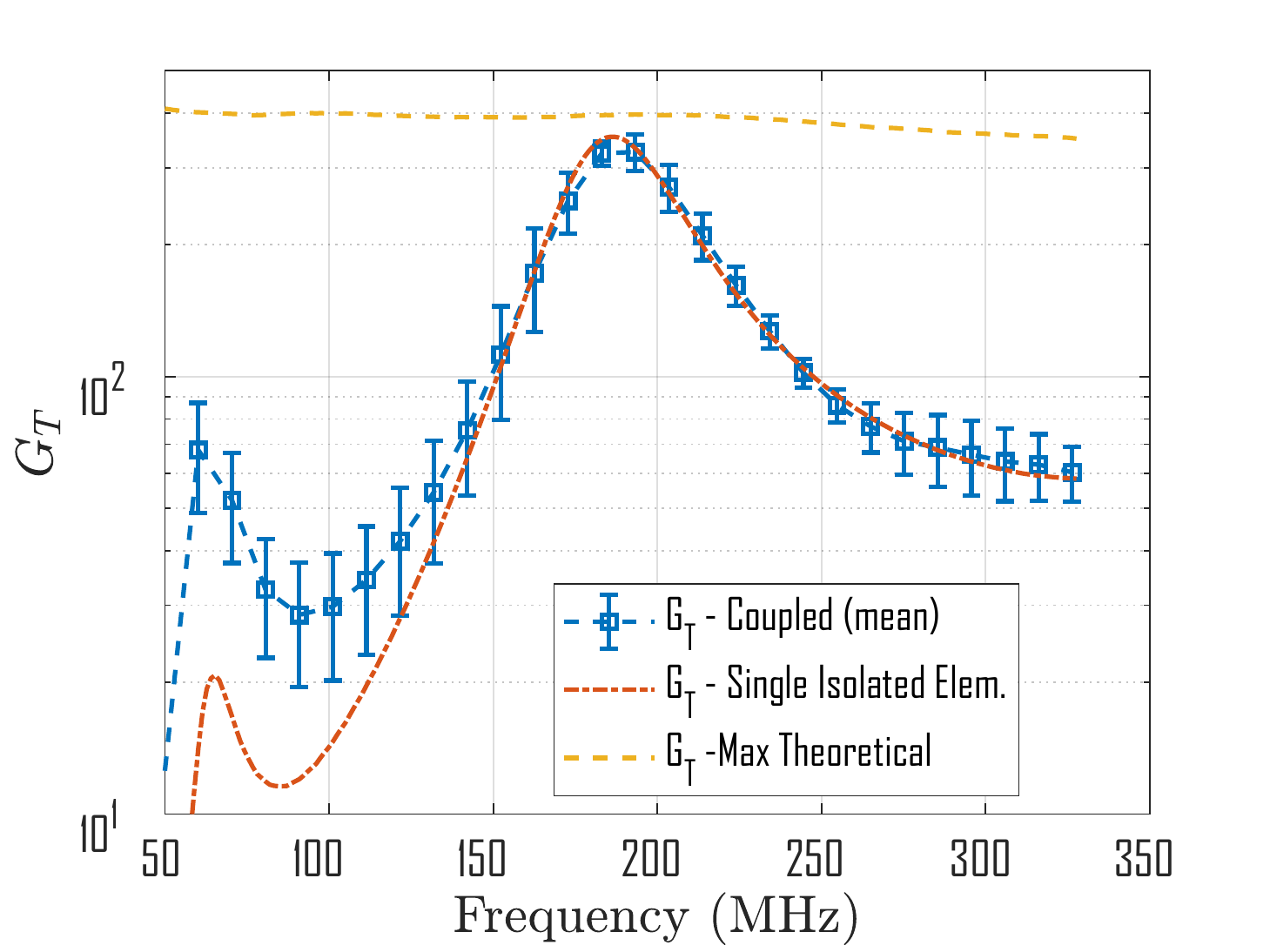}
	\caption{Comparison of the transducer gain achievable by the MWA LNA. The theoretical maximum transducer gain (dashed curve) of a single element was obtained by placing conjugately matched load at all frequencies at the input of the LNA. The data points represent the mean and standard deviation of the tile's $G_T$ obtained over all 197 optimal pointing angles. For comparison, $G_T$ of a single isolated element (dot-dash curve) is shown. \label{fig:MWA_GA_comparison}}
\end{figure}

The most interesting result that emerged from this calculation is the reduction of $T'_{\mathrm{rcv}}(\nu)$ over the single element at lower frequencies. As shown in Fig. \ref{fig:pint_comparison}, the lowering of $T_{\mathrm{rcv}}$ can be attributed to increasing $G_T$ as the total internal noise power delivered to the load in the array environment is fairly similar to the single isolated element case. To verify that the results presented remain physically valid, $G_{T}$ was recalculated for a single isolated element given that the antenna was conjugately matched at all frequencies to the LNA's input impedance. This calculation sets the absolute upper limit which the $G_{T}$ of the array must not exceed as it would imply the source is delivering more power than the available source power ($kT_{0}$). This maximum value was compared to the mean and standard deviation of the array's $G_{T}$ over all 197 pointing angles in Fig. \ref{fig:MWA_GA_comparison} and showed that the calculated array $G_{T}$ remains physically valid. 

\begin{figure}[]
	\centering
	\includegraphics[width=\linewidth]{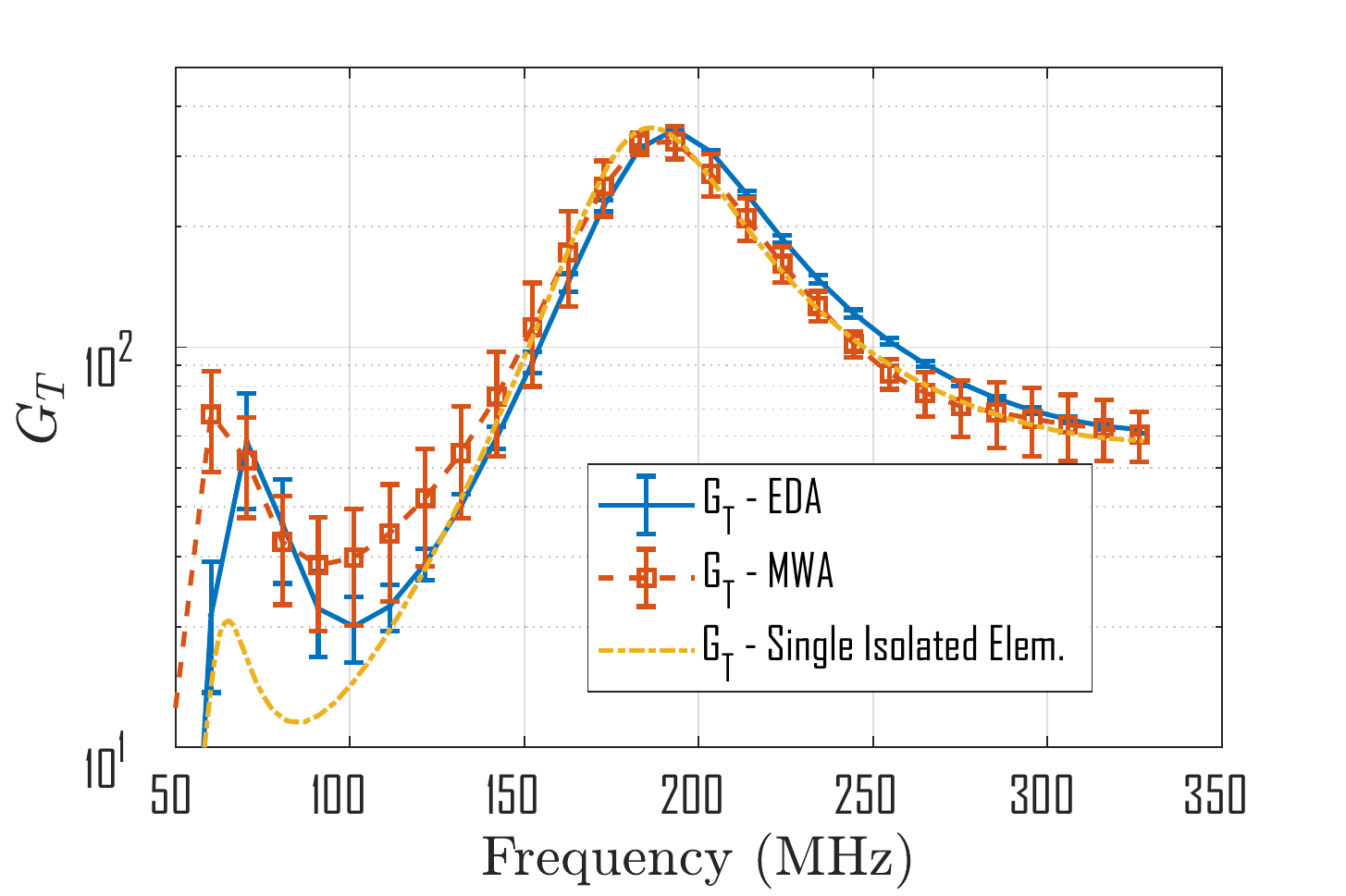}
	\caption{Comparison of mean and standard deviation of calculated transducer gain of the MWA and EDA over 197 optimal MWA pointings. The $G_T$ of a single isolated element is represented by the dashed curve for reference.\label{fig:GA_comparison}}
\end{figure}

Similar effects have been observed in the EDA results seen in Fig. \ref{fig:EDA_comparison}. However, the reduction of $T'_{\mathrm{rcv}}(\nu)$ \added{when compared to a single element} was not as drastic as the transducer gain did not increase as much as the MWA tile. Standard deviations in Fig. \ref{fig:GA_comparison} clearly show that in general MWA has a higher transducer gain when compared to the EDA, however the $G_{T}$ of an MWA tile varies over pointing angles more than the EDA. 

\begin{figure}[]
	\centering
	\includegraphics[width=\linewidth]{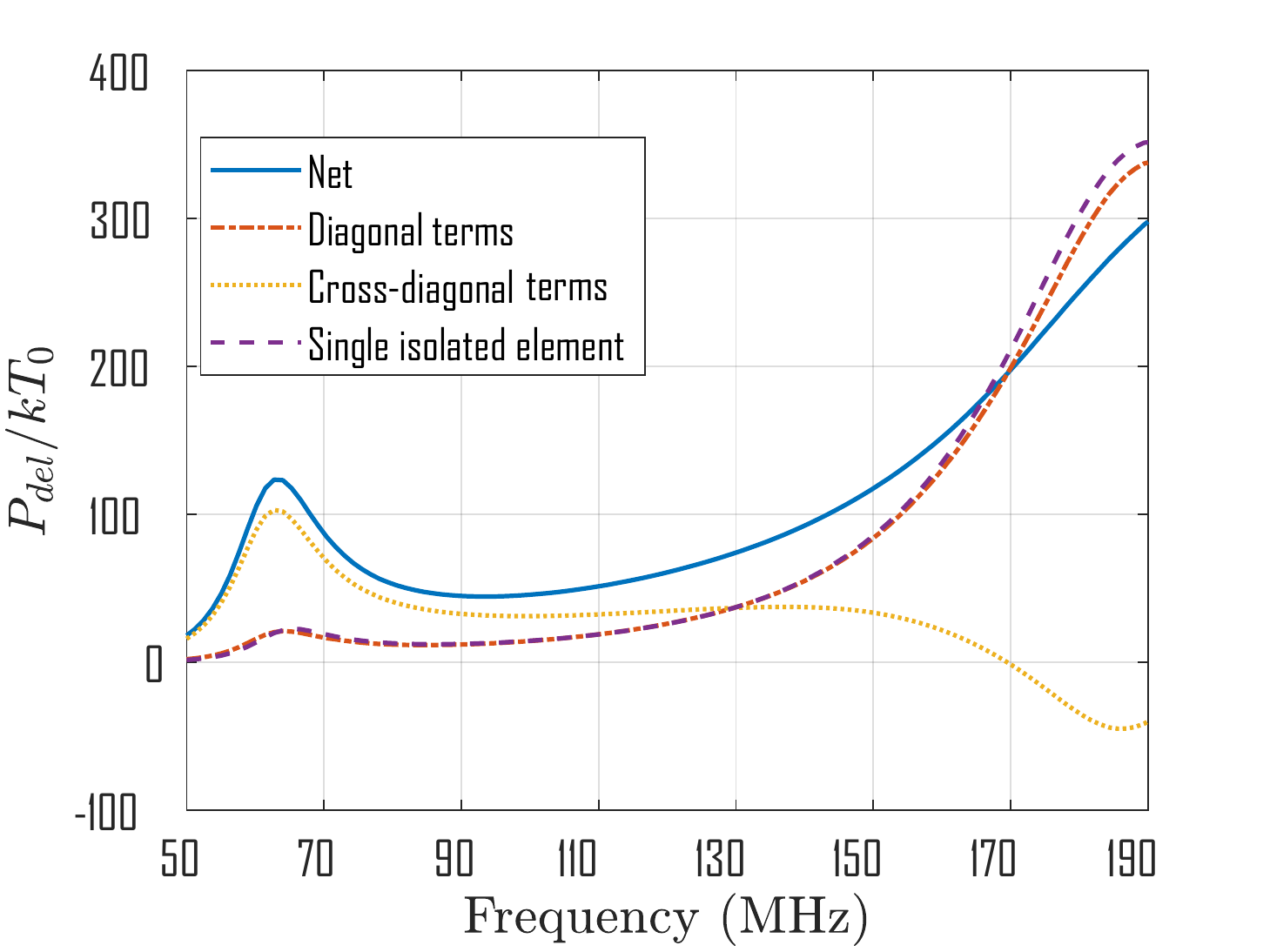}
	\caption{\added{External noise power due to homogeneous sky delivered to the $Z_0$ matched load at the output of MWA tile normalized to $kT_0$. The dashed curve represents the total delivered power due to external sources for a single isolated element, the dotted curve represents the additional delivered power to the MWA tile due to mutual coupling (cross terms of \eqref{eqn:P_ext_randa}) while the dot-dash curve represents external noise power that is directly delivered to the array (diagonal terms of \eqref{eqn:P_ext_randa}). The solid curve is the net sum of both the additional and direct power delivered to the MWA tile. The negative value represents power loss due to destructive interference of the noise wave due to mutual coupling.}\label{fig:Pdel_MWA}}
\end{figure}

\begin{figure}[]
	\centering
	\includegraphics[width=\linewidth]{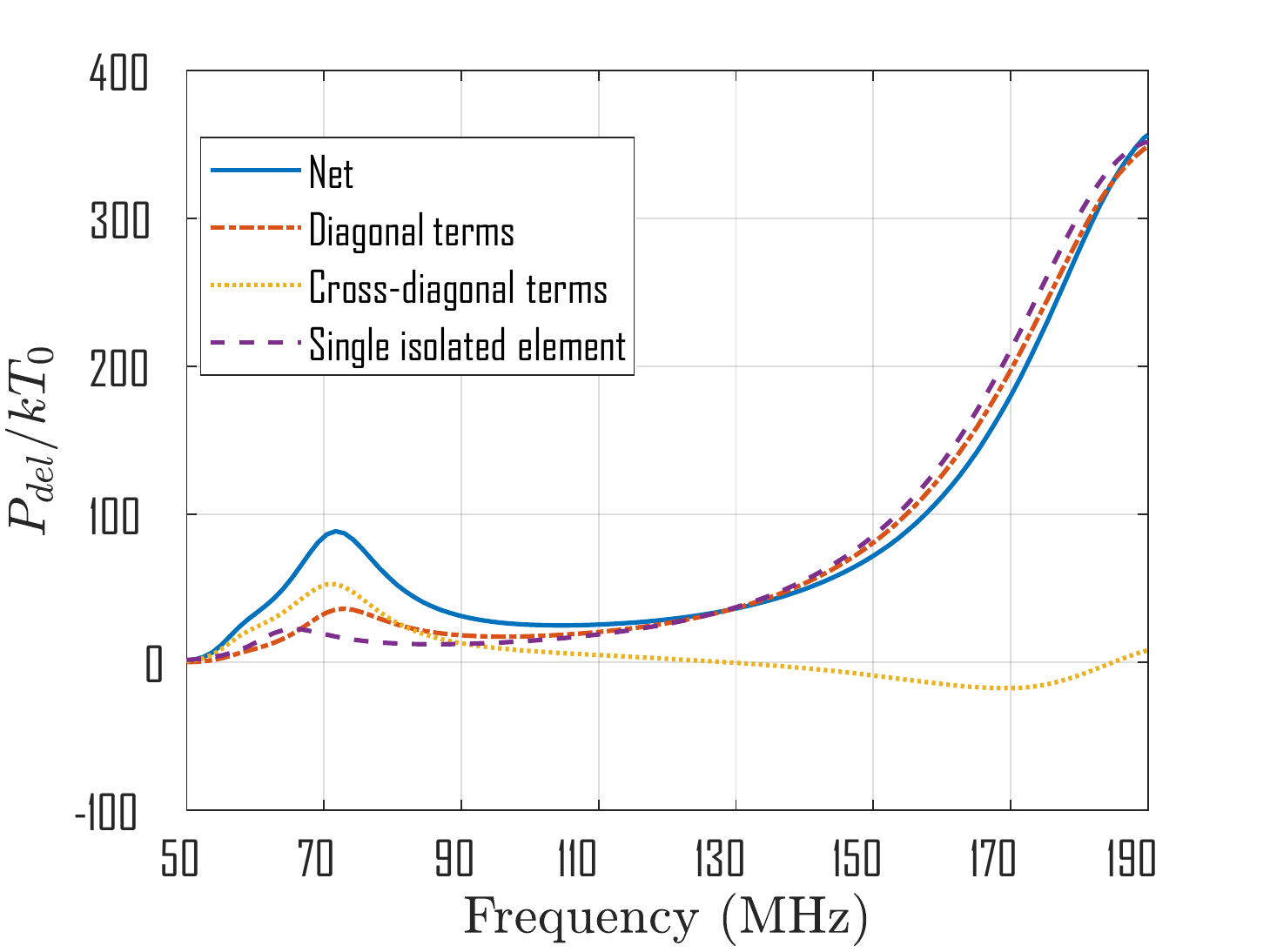}
	\caption{\added{External noise power due to homogeneous sky delivered to the $Z_0$ matched load at the output of the EDA normalized to $kT_0$. The dashed curve represents the total delivered power due to external sources for a single isolated element, the dotted curve represents the additional delivered power to the EDA due to mutual coupling (cross terms of \eqref{eqn:P_ext_randa}) while the dot-dash curve represents external noise power that is directly delivered to the array (diagonal terms of \eqref{eqn:P_ext_randa}). The solid curve is the net sum of both the additional and direct power delivered to the EDA. The negative value represents power loss due to destructive interference of the noise wave due to mutual coupling.}\label{fig:Pdel_EDA}}
\end{figure}

\added{The mechanism causing the increasing of $G_{T}$ can be investigated} by plotting the delivered power to the array due to external sources alone as shown in Fig. \ref{fig:Pdel_MWA} and \ref{fig:Pdel_EDA}. \added{It can be observed} that additional power is delivered to the array due to mutual coupling at 50 to 170 \si{\mega\hertz}. It can be clearly seen that for the MWA, the reduction in $T_{\mathrm{rcv}}$ is due to more external noise power (signal of interest) being delivered to the array. In contrast, the EDA has less coupling, hence less additional power is delivered to the array which leads to less reduction in $T_{\mathrm{rcv}}$. 

The higher levels of power delivered due to mutual coupling (cross terms) makes the array sensitive to complex weightings applied by the beamformer at the output. This result is consistent with the trend observed in Fig. \ref{fig:GA_comparison} whereby, the MWA experiences larger changes to $G_{T}$ with changing pointing angles. Such high level of coupling can be explained by the physical layout of the elements. The element spacing within an MWA tile is $1.1$ \si{\meter} from centre to centre whereas for the EDA, the average element spacing is $\approx1.5$ \si{\meter}. 

Based on these results, it can be reasoned that the reduction of $T_{\mathrm{rcv}}$ is possible by optimizing the element layout (increased coupling) without having to optimize the LNA over the entire frequency band. Meaning, the LNA could be optimized to cover the mid to high frequency band whereas, the array layout could be optimized to take advantage of the effects of mutual coupling to improve the $T_{\mathrm{rcv}}$ at lower frequencies (50 - 140 \si{\mega\hertz}). 

Layout optimization only works if the dominant contribution is due to external noise. That is to say, the if the internal noise is not fluctuating much as a function of complex weightings (see Fig. \ref{fig:pint_comparison}), \added{then increasing} the amount of external noise through mutual coupling is beneficial for lowering $T_{\mathrm{rcv}}$. In the domain where internal noise dominates, increased mutual coupling is not desirable as this will lead to an increase in $T_{\mathrm{rcv}}$. The only way to decrease $T_{\mathrm{rcv}}$ in this case without changing the antenna design is to optimize the LNA. \added{This reasoning comes with a caveat that it only applies to the MWA dipole design. Other antenna designs were not analyzed which could potentially lead to a different conclusion found here.}

\section{Conclusion}
\label{sec:conlc}
This paper presents a power wave based framework for analyzing the $T_{\mathrm{rcv}}$ of an aperture array which includes the effects of mutual coupling. Using a combination of measured noise parameters and simulated $S$-parameters of the MWA tile and the EDA to calculate the receiver noise temperature. \added{The calculated $T_{\mathrm{rcv}}$ obtained using the proposed PWF was compared} with measured $T_{\mathrm{rcv}}$ obtained via astronomical observations and was found to be in good agreement between the two for both the MWA tile and the EDA. It was observed that due to higher mutual coupling in the $50-140~\si{\mega\hertz}$ region, the MWA has a lower receiver noise when compared to the EDA. The decrease in $T_{\mathrm{rcv}}$ was due to the increase in transducer gain.

The increased $G_T$ at lower frequencies was due to the additional external noise power delivered to the array via coupling. This improvement was seen for both the MWA tile and the EDA but since the MWA tile has higher coupling, the $T_{\mathrm{rcv}}$ was lower than the EDA. In addition, higher fluctuation in $G_T$ as a function of pointing angles with higher levels of coupling. For this reason, mutual coupling could either be a hindrance or aid when it comes to reducing $T_{\mathrm{rcv}}$ depending on whether the internal or external noise dominates the overall contribution. 

Additionally, it was demonstrated that the PWF is able to make use of embedded element patterns to calculate the efficiency of the array without the need for re-simulation. In conclusion, the PWF presented in \cite{954781} is a general method suited to compute receiver noise temperature for multiport devices and extendable to include phased arrays as demonstrated in this paper. This framework can be utilized to optimize and characterize future generation telescopes such as the Square Kilometre Array \cite{5136190}. 

\appendices
\section{Derivation of $\mathbf{M}$ Matrix}
\label{sec:mismatch}
The standard $S$-parameter representation of incident and reflected wave are as shown below.
\begin{align}
\label{eqn:a}
\mathbf{a} &= \mathbf{S}_{\mathrm{LNA}}\mathbf{b} + \mathbf{n}\\
\label{eqn:b}
\mathbf{b} &= \mathbf{S}_{\mathrm{load}}\mathbf{a}
\end{align}
where $\mathbf{a}$ is a vector containing the outgoing wave from the the input and output ports of multiport amplifier indicated by the red arrows in Fig. \ref{fig:port_numbering}, $\mathbf{n}$ is a vector containing the noise waves $c_1$ and $c_2$, $\mathbf{b}$ is the vector containing the reflected wave due to the attached load at the input and output ports of the network, $\mathbf{S}_{\mathrm{LNA}}$ and $\mathbf{S}_{\mathrm{load}}$ are the $S$-parameters of the network and loads respectively.

The noise wave vector $\mathbf{n}$ appears in \eqref{eqn:a} to represent noise wave originating from the network. By substituting \eqref{eqn:b} into \eqref{eqn:a} and solving for $\mathbf{a}$ yields,
\begin{align}
\label{eqn:M_derivation}
\mathbf{a} &= \left[1-\mathbf{S}_{\mathrm{LNA}}\mathbf{S}_{\mathrm{load}}\right]^{-1}\mathbf{n}\\
\mathbf{a} &= \mathbf{M}\mathbf{n}.
\end{align}

The outgoing power due to internal noise alone is simply given by 
\begin{align}
\label{eqn:outgoing_pwr}
\mathbf{A}^{\mathrm{out}}_{\mathrm{int}} &= \mathbf{a}\mathbf{a}^\dagger\\
&= \mathbf{M}\mathbf{\hat{N}}\mathbf{M}^\dagger
\end{align}
where $\hat{\mathbf{n}\mathbf{n}}^\dagger = \mathbf{\hat{N}}$.

If desired, $\mathbf{n}$ can be shifted to \eqref{eqn:b} and this represents noise originating from the loads (external noise). Following the exact derivation shown above will yield \eqref{eqn:P_ext_randa}. Solving for $\mathbf{b}$ and repeating steps above on the other hand, will produce expression seen in \eqref{eqn:p_inc_ext} and \eqref{eqn:M_dash}.

\section{Derivation of $\mathbf{l_{\mathrm{p}}}$}
\label{sec:norm_pat}
The quantity $\mathbf{l}_{\mathrm{p}}$ is defined as follows
\begin{align}
\label{eqn:v_lna}
V_{\mathrm{LNA}} &= \frac{Z_{L}}{Z_{L} + Z_{tx}}\mathbf{l}_{\mathrm{eff}}\mathbf{E}^{\mathrm{inc}}\\
 &= \mathbf{l}_{\mathrm{p}}\mathbf{E}^{\mathrm{inc}}
\end{align}
where the effective length ($\mathbf{l}_{\mathrm{eff}}$) is given by \cite{warnick_maaskant_ivashina_davidson_jeffs_2018}
\begin{align}
\label{eqn:l_eff}
\mathbf{l}_{\mathrm{eff}} = -j\frac{4\pi}{\omega\mu_{0}I_{tx}}\bar{\mathbf{E}}
\end{align}
where $V_{\mathrm{LNA}}$ is the voltage dropped across the input of the LNA, $\mathbf{E}^{\mathrm{inc}}$ is the incident plane wave, $Z_{L}$ and $Z_{tx}$ are the impedance of the load and antenna under transmit condition respectively, $\omega$ is the angular frequency, $\mu_{0}$ is the permeability of free space, $I_{tx}$ is the port current under transmit condition and $\bar{\mathbf{E}} = \left[E_{\hat{\theta}}^{tx},E_{\hat{\phi}}^{tx}\right]$ is embedded element radiation pattern.

Realizing that $I_{tx}$ is simply
\begin{align}
I_{tx} = \frac{V_{tx}}{Z_{L} + Z_{tx}}
\label{eqn:i_tx}
\end{align}
and by substituting \eqref{eqn:i_tx} into \eqref{eqn:l_eff} to obtain
\begin{align}
\centering
\mathbf{l}_{\mathrm{p}} =-j\frac{4\pi}{\omega\mu_{0}}\frac{Z_{L}}{V_{tx}}[E_{\hat{\theta}}^{tx},E_{\hat{\phi}}^{tx}]^T
\label{eqn:eff_len}
\end{align}
where $V_{tx}$ is the source voltage applied in simulation which produces the corresponding $\bar{\mathbf{E}}$ and $Z_{tx}$ is the impedance of the antenna given that all other surrounding elements are terminated with a load impedance $Z_{L}$. 

\section*{ACKNOWLEDGEMENT}
The author would like to thank Randall Wayth for useful feedback on this manuscript. The author acknowledges the contribution of an Australian Government Research Training Program Scholarship in supporting this research. The International Centre for Radio Astronomy Research (ICRAR) is a Joint Venture of Curtin University and The University of Western Australia, funded by the Western Australian State government. The MWA Phase II upgrade project was supported by Australian Research Council LIEF grant LE160100031 and the Dunlap Institute for Astronomy and Astrophysics at the University of Toronto. This scientific work makes use of the Murchison Radio-astronomy Observatory, operated by CSIRO. We acknowledge the Wajarri Yamatji people as the traditional owners of the Observatory site. Support for the operation of the MWA is provided by the Australian Government (NCRIS), under a contract to Curtin University administered by Astronomy Australia Limited. We acknowledge the Pawsey Supercomputing Centre which is supported by the Western Australian and Australian Governments.

\bibliographystyle{IEEEtran}
\bibliography{MWA_array_temp_reference}

\end{document}